\documentclass[aps,prd,twocolumn,superscriptaddress,showpacs]{revtex4}

\usepackage{epsfig,epsf}
\usepackage{amsmath}
\usepackage{amsthm}
\usepackage{amsfonts}
\usepackage{amssymb}
\usepackage{dsfont}
\usepackage{multirow}
\usepackage{appendix}
\usepackage{slashed}
\usepackage[active]{srcltx}
\usepackage{psfrag}
\usepackage{graphicx}
\usepackage{multirow}
\usepackage{subfigure}
\input epsf

\usepackage{subfigure}
\begin{document}

\title{Decuplet baryons in a hot medium}
\date{\today}
\author{K. Azizi, G. Bozk{\i}r  \\
\textit{Department of Physics, Do\v{g}u\c{s} University, Ac{\i}badem-Kad{\i}k\"oy, \\ 34722 Istanbul, Turkey}}

\begin{abstract}

The thermal properties of the light decuplet baryons  are investigated in the framework of the thermal QCD sum rules. 
In particular, the behavior of the mass and residue of the $\Delta$, $\Sigma^{*}$, $\Xi^{*}$ and $\Omega$ baryons with respect to  temperature are analyzed taking into account 
the additional operators coming up in the Wilson expansion at finite temperature. It is found that the mass and residue of these particles  remain overall unaffected up to $T\simeq150~MeV$ but, 
beyond this point, they start to diminish, considerably.  

\end{abstract}

\maketitle



%

\section{Introduction}

%
Over the past decade,  the in-medium investigation of fundamental parameters of hadrons  such as their mass, decay constant, widths and strong coupling constants among various hadrons has become
one of the hot topics in hadron physics. Such investigation can help us not only better analyze the  hot and dense QCD matter produced by heavy ion collision experiments, bu also
can help us  get valuable knowledge on the perturbative and nonperturbative natures of QCD.   
  Investigations of  parameters of different members of light baryons can also provide us with useful information on  the structure of dense astrophysical objects like neutron stars 
since at the center of the neutron stars there can be intensively produced
 the strange members of octet baryons as well as the decuplet baryons beside the nucleons.  

The theoretical studies on the hadronic parameters at finite temperature have mainly been devoted to mesons. There are a few studies dedicated to the studies of the 
light  baryons in hot medium. It is well known that the investigation of parameters of both the decuplet and octet baryons at finite temperature is very important in
  understanding of the SU(3) flavor chiral symmetry breaking. In Ref. \cite{Bedaque},  the temperature dependence of the pole mass of the octet and decuplet baryons is investigated using 
the  chiral perturbation theory. 
In Ref. \cite{Rincon}, the baryon masses at finite temperature has been investigated in both the octet and decuplet representations for the Nambu-Jona-Lasinio (NJL) 
and Polyakov-Nambu-Jona-Lasinio (PNJL) models. It has been obtained that the baryon masses decrease by increasing the temperature and there is a strong dependence on the melting
 or deconfinement temperature determined by the flavor content of the baryons. In Ref. \cite{Xu}, the mass of the decuplet baryons at finite temperature are investigated using thermal QCD sum rules.
 In this paper, the authors find that  below $T\leq0.11~GeV$ the masses show very little dependence on the temperature, but decrease with increasing temperature above this point. 
The self energy of the $\Delta$-baryon is also investigated at finite temperature and density using the real time formalism of thermal field theory in Ref.  \cite{Ghosh}.

The aim of this article is to extend our previous studies on the thermal properties of nucleon and hyperons  \cite{Azizi1, Azizi2}  and  evaluate the behavior of the  mass and residue of
 the decuplet $\Delta$, $\Sigma^{*}$, $\Xi^{*}$ and $\Omega^{-}$ baryons with respect to temperature employing the thermal QCD sum rule method. 
This method is the extended version of the vacuum QCD sum rule \cite{Shifman} to finite temperature and first introduced by Bochkarev and Shaposhnikov \cite{Bochkarev}. 
This method has two new aspects compared to the case of zero temperature:
 the Lorentz invariance is broken at finite temperature with the choice of reference frames, to restore of which the four-velocity vector of the medium is introduced; and 
  some new operators appear in the operator product expansion (OPE) at finite temperature and the vacuum condensates are replaced by their thermal expectation values. 
In our calculations, we use the thermal quark propagators containing new non-perturbative contributions coming up in the Wilson expansion at finite temperature.  We use the expressions of the
 temperature- dependent quark and gluon condensates as well as the temperature-dependent fermionic and gluonic parts of the energy density calculated via lattice QCD. 
We find the temperature-dependent expression of the continuum threshold in decuplet channel using the obtained sum rules for the masses and residues.

This study is organized as follows. In section \ref{sec:DA}, the thermal QCD sum rules for the mass and residue of  $\Delta$, $\Sigma^{*}$, $\Xi^{*}$ and $\Omega^{-}$ decuplet baryons
 are derived. In section \ref{sec:results},  the numerical analyses of the obtained sum rules for decuplet baryons as well as comparison of the results with those existing in the literature are presented.

%

\section{Thermal QCD sum rules for the mass and residue of  decuplet baryons}

\label{sec:DA} %

In this section, thermal QCD sum rules for the mass and residue of $\Delta$, $\Sigma^{*}$, $\Xi^{*}$ and $\Omega$ decuplet baryons are derived. For this purpose,
 we choose the following two-point thermal correlation function:
\begin{equation}\label{eqn1}
\Pi_{\mu\nu}(p,T)=i \int d^{4}x e^{ip\cdot x}~ Tr\{\rho {\cal
T}[J^{D}_{\mu}(x)\bar{J}^{D}_{\nu}(0)]\}, \\
\end{equation}
where $J^{D}_{\mu}(x)$ is the interpolating current of \textbf{D} decuplet baryon, $\rho= e^{-\beta H}/Tr e^{-\beta H}$ is the thermal density
matrix of QCD with $H$ being  the QCD Hamiltonian,  $T=1/\beta$ is temperature and $\cal T$ indicates the time ordering operator.
%
For $J^{D}_{\mu}(x)$ interpolation current, we use the following expression in a compact form (see for instance \cite{Aliev:2010dw,Aliev:2009pd,Lee}):

\begin{eqnarray}
J^{\textbf{D}}_{\mu}(x) &=&N \epsilon_{abc}\
\Bigg[\Big( q_1^{T,a}(x)C\gamma _{\mu}q_2^{b}(x)\Big) q_3^{c}(x) \notag \\
&+& \Big( q_2^{T,a}(x)C\gamma _{\mu}q_3^{b}(x)\Big) q_1^{c}(x) \notag \\
&+& \Big( q_3^{T,a}(x)C\gamma _{\mu}q_1^{b}(x)\Big) q_2^{c}(x) \Bigg],
\label{currents1}
\end{eqnarray}
where $a, b, c$ are color indices and $C$ denotes the charge conjugation operator. The values
of normalization constant $N$ and the $q_1, q_2$ and $q_3$ quarks for the considered light decuplet baryons are given in table \ref{table1}.





\begin{table}[h]

\renewcommand{\arraystretch}{1.3}
\addtolength{\arraycolsep}{-0.5pt}
\small
$$
\begin{array}{|l|c|c|c|c|}
\hline \hline
 & N & q_1 & q_2 & q_3 \\  \hline
\Delta^0 & \sqrt{1/3}  & d & d & u  \\
\Sigma^{*0} & \sqrt{2/3}  & u & d & s  \\
\Xi^{*0} & \sqrt{1/3}  & s & s & u  \\
\Omega^{-}    &       1/3   & s & s & s  \\
\hline \hline
\end{array}
$$
\caption{The values of $N$ and the quark flavors $q_1$, $q_2$
and $q_3$ for $\Delta^0$, $\Sigma^{*0}$, $\Xi^{*0}$ and $\Omega^{-}$ decuplet baryons.}
\renewcommand{\arraystretch}{1}

\addtolength{\arraycolsep}{-1.0pt}\label{table1}

\end{table}

To construct the thermal sum rules for the considered decuplet baryons, the aforesaid correlation function is computed both in hadronic  and OPE representations. 
Matching then the  coefficients of sufficient structures from the two representations, through a dispersion relation, the sum rules for the mass and residue of decuplet baryons are obtained. 
In order to suppress the contributions of the higher resonances and continuum, a Borel transformation as well as  continuum subtraction are performed. 

The hadronic side of the correlation function is obtained by inserting a complete set of intermediate state with spin $s$ into Eq. (\ref{eqn1}). After performing the integral over four-$x$, we get

\begin{eqnarray}\label{phepi}
\Pi_{\mu\nu}^{Had}(p,T)&=&-\frac{{\langle}0|J^{D}_{\mu}(0)|D(p,s){\rangle}_T
{\langle}D(p, s)|J^{D\dag}_{\nu}(0)|0{\rangle}_T}{p^{2}-m_{D}^{2}(T)} \notag \\
&+&...,
\end{eqnarray}
where  $...$ denotes the contributions of the higher states and continuum and $m_{D}(T)$ is the temperature-dependent mass of decuplet baryons. The matrix element  ${\langle}0|J^{D}_{\mu}(0)|D(p,s){\rangle}_T$ is defined as 

\begin{eqnarray}\label{intcur}
{\langle}0|J^{D}_{\mu}(0)|D(p,s){\rangle}_T&=&\lambda_{D}(T)u_{\mu}(p,s) ,
\end{eqnarray}
where $u_{\mu}(p,s)$ is the Rarita-Schwinger spinor and $\lambda_{D}(T)$ is the  temperature-dependent  residue of  $D$  baryon.
 We shall remark that, $J^{D}_{\mu}$ current couples to  not only the  spin-$3/2$ but also the spin-$1/2$ states. To get only the contributions of the decuplet baryons,
 we need to remove the unwanted contribution coming from the spin-$1/2$ states. To this end we employ the following procedures.
 The matrix element of $J^{D}_{\mu}$ between the spin-$1/2$ and vacuum states can be written as

\begin{eqnarray}\label{intcur1}
{\langle}0|J^{D}_{\mu}|\frac{1}{2}(p){\rangle}_T&=&\Big(A p_{\mu}+B \gamma_{\mu}\Big)u(p),
\end{eqnarray}
where $A$ and $B$ are constants. By multiplication of both sides of Eq. (\ref{intcur1}) with $\gamma_{\mu}$, 
and  taking into account the condition $J^{D}_{\mu}\gamma^{\mu}=0$, one immediately finds $A$ in terms of  $B$, i.e.

\begin{eqnarray}\label{intcur2}
{\langle}0|J^{D}_{\mu}(0)|\frac{1}{2}(p){\rangle}_T&=&B\Big(-\frac{4}{m_{\frac{1}{2}}} p_{\mu}+ \gamma_{\mu}\Big)u(p).
\end{eqnarray}
From this equation we notice that the unwanted contributions coming from the spin-$1/2$ states are proportional to either $p_{\mu}$ or $\gamma_{\mu}$. 
To eliminate these contributions,  we order the Dirac matrices as $\gamma_{\mu}\!\not\!{p}\gamma_{\nu}$ and  set to zero the terms with $\gamma_{\mu}$ 
in the beginning and $\gamma_{\nu}$ at the end and those  proportional to $p_{\mu}$ and  $p_{\nu}$.


Finally, by inserting Eq. (\ref{intcur}) into Eq. (\ref{phepi}) and summing over the spin-$3/2$ states, the correlation function in hadronic side in the Borel scheme is obtained as

\begin{eqnarray}
&&\hat{B}\Pi_{\mu\nu}^{Had}(p,T)=-\Bigg[\lambda_D^{2}(T)e^{-m_{D}^{2}(T)/M^2}\Bigg]\!\not\!{p}g_{\mu\nu}\notag \\
&-&\Bigg[\lambda_D^{2}(T)m_{D}(T)e^{-m_{D}^{2}(T)/M^2}\Bigg]g_{\mu\nu}
+\textsf{other structures}\notag \\
&&\textsf {including the  four velocity vector of the medium},
\end{eqnarray}
 where $M^2$  is the Borel parameter to be fixed in next section.
%

On the other hand, the OPE side of the thermal correlation function is calculated in terms of  the quark-gluon degrees of freedom in deep Euclidean region. 
By inserting the explicit forms of the interpolating currents into the correlation function given in Eq. (\ref{eqn1}) and contracting out all quark pairs via Wick's theorem, we obtain 
the OPE side of thermal correlation function for the considered decuplet baryons in terms of the thermal light quarks propagators:
\begin{widetext}
\begin{eqnarray}\label{corre1}
\Pi_{\mu\nu}^{OPE,\Delta}(p,T) &=& \frac{i}{3}\epsilon_{abc}\epsilon_{a'b'c'}\int d^4 x e^{ipx} \Bigg\langle \Bigg\{2S^{ca'}_{d}(x)\gamma_{\nu}S'^{ab'}_{d}(x)\gamma_{\mu}S^{bc'}_{u}(x)-2S^{cb'}_{d}(x)\gamma_{\nu}S'^{aa'}_{d}(x)\gamma_{\mu}S^{bc'}_{u}(x)\nonumber\\
&+&4S^{cb'}_{d}(x)\gamma_{\nu}S'^{ba'}_{u}(x)\gamma_{\mu}S^{ac'}_{d}(x)+2S^{ca'}_{u}(x)\gamma_{\nu}S'^{ab'}_{d}(x)\gamma_{\mu}S^{bc'}_{d}(x)\nonumber\\
&-&2S^{ca'}_{u}(x)\gamma_{\nu}S'^{bb'}_{d}(x)\gamma_{\mu}S^{ac'}_{d}(x)-S^{cc'}_{u}(x)Tr\Bigg[S^{ba'}_{d}(x)\gamma_{\nu}S'^{ab'}_{d}(x)\gamma_{\mu}\Bigg]\nonumber\\
&+&S^{cc'}_{u}(x)Tr\Bigg[S^{bb'}_{d}(x)\gamma_{\nu}S'^{aa'}_{d}(x)\gamma_{\mu}\Bigg]-4S^{cc'}_{d}(x)Tr\Bigg[S^{ba'}_{u}(x)\gamma_{\nu}S'^{ab'}_{d}(x)\gamma_{\mu}\Bigg]\Bigg\} \Bigg\rangle_{T},
\end{eqnarray}
\begin{eqnarray}\label{corre2}
\Pi_{\mu\nu}^{OPE,\Sigma^{*}}(p,T) &=& -\frac{2i}{3}\epsilon_{abc}\epsilon_{a'b'c'}\int d^4 x e^{ipx}\Bigg\langle \Bigg\{S^{ca'}_{d}(x)\gamma_{\nu}S'^{bb'}_{u}(x)\gamma_{\mu}S^{ac'}_{s}(x)\nonumber\\
&+&S^{cb'}_{d}(x)\gamma_{\nu}S'^{aa'}_{s}(x)\gamma_{\mu}S^{bc'}_{u}(x)+S^{ca'}_{s}(x)\gamma_{\nu}S'^{bb'}_{d}(x)\gamma_{\mu}S^{ac'}_{u}(x)\nonumber\\
&+&S^{cb'}_{s}(x)\gamma_{\nu}S'^{aa'}_{u}(x)\gamma_{\mu}S^{bc'}_{d}(x)+S^{ca'}_{u}(x)\gamma_{\nu}S'^{bb'}_{s}(x)\gamma_{\mu}S^{ac'}_{d}(x)\nonumber\\
&+&S^{cb'}_{u}(x)\gamma_{\nu}S'^{aa'}_{d}(x)\gamma_{\mu}S^{bc'}_{s}(x)+S^{cc'}_{s}(x)Tr\Bigg[S^{ba'}_{d}(x)\gamma_{\nu}S'^{ab'}_{u}(x)\gamma_{\mu}\Bigg]\nonumber\\
&+&S^{cc'}_{u}(x)Tr\Bigg[S^{ba'}_{s}(x)\gamma_{\nu}S'^{ab'}_{d}(x)\gamma_{\mu}\Bigg]+S^{cc'}_{d}(x)Tr\Bigg[S^{ba'}_{u}(x)\gamma_{\nu}S'^{ab'}_{s}(x)\gamma_{\mu}\Bigg]\Bigg\} \Bigg\rangle_{T},
\end{eqnarray}
\begin{eqnarray}\label{corre3}
\Pi_{\mu\nu}^{OPE,\Xi^{*}}(p,T) &=& \frac{i}{3}\epsilon_{abc}\epsilon_{a'b'c'}\int d^4 x e^{ipx}\Bigg\langle \Bigg\{2S^{ca'}_{s}(x)\gamma_{\nu}S'^{ab'}_{s}(x)\gamma_{\mu}S^{bc'}_{u}(x)\nonumber\\
&-&2S^{cb'}_{s}(x)\gamma_{\nu}S'^{aa'}_{s}(x)\gamma_{\mu}S^{bc'}_{u}(x)+4S^{cb'}_{s}(x)\gamma_{\nu}S'^{ba'}_{u}(x)\gamma_{\mu}S^{ac'}_{s}(x)\nonumber\\
&+&2S^{ca'}_{u}(x)\gamma_{\nu}S'^{ab'}_{s}(x)\gamma_{\mu}S^{bc'}_{s}(x)-2S^{ca'}_{u}(x)\gamma_{\nu}S'^{bb'}_{s}(x)\gamma_{\mu}S^{ac'}_{s}(x)\nonumber\\
&-&S^{cc'}_{u}(x)Tr\Bigg[S^{ba'}_{s}(x)\gamma_{\nu}S'^{ab'}_{s}(x)\gamma_{\mu}\Bigg]+S^{cc'}_{u}(x)Tr\Bigg[S^{bb'}_{s}(x)\gamma_{\nu}S'^{aa'}_{s}(x)\gamma_{\mu}\Bigg]\nonumber\\
&-&4S^{cc'}_{s}(x)Tr\Bigg[S^{ba'}_{u}(x)\gamma_{\nu}S'^{ab'}_{s}(x)\gamma_{\mu}\Bigg]\Bigg\} \Bigg\rangle_{T}, 
\end{eqnarray}
\begin{eqnarray}\label{corre4}
\Pi_{\mu\nu}^{OPE,\Omega^{-}}(p,T) &=& \epsilon_{abc}\epsilon_{a'b'c'}\int d^4 x e^{ipx}\Bigg\langle \Bigg\{S^{ca'}_{s}(x)\gamma_{\nu}S'^{ab'}_{s}(x)\gamma_{\mu}S^{bc'}_{s}(x)\nonumber\\
&-&S^{ca'}_{s}(x)\gamma_{\nu}S'^{bb'}_{s}(x)\gamma_{\mu}S^{ac'}_{s}(x)-S^{cb'}_{s}(x)\gamma_{\nu}S'^{aa'}_{s}(x)\gamma_{\mu}S^{bc'}_{s}(x)\nonumber\\
&+&S^{cb'}_{s}(x)\gamma_{\nu}S'^{ba'}_{s}(x)\gamma_{\mu}S^{ac'}_{s}(x)-S^{cc'}_{s}(x)Tr\Bigg[S^{ba'}_{s}(x)\gamma_{\nu}S'^{ab'}_{s}(x)\gamma_{\mu}\Bigg]\nonumber\\
&+&S^{cc'}_{s}(x)Tr\Bigg[S^{bb'}_{s}(x)\gamma_{\nu}S'^{aa'}_{s}(x)\gamma_{\mu}\Bigg]\Bigg\} \Bigg\rangle_{T}, 
\end{eqnarray}
where $S'=CS^TC$. 
\end{widetext}
Note that the same structures as the hadronic side enter to the OPE side. The unwanted contributions are removed with the same procedures as
the hadronic side of the correlation function.

To proceed, we need the thermal light quark propagator, whose expression (expanded in terms of different operators having different mass dimensions)   in $x$ space is given as (see also \cite{Azizi2, Veliev})
\begin{eqnarray}\label{lightquarkpropagator}
S_{q}^{ij}(x)&=& i\frac{\!\not\!{x}}{ 2\pi^2 x^4}\delta_{ij}-\frac{m_q}{4\pi^2 x^2}\delta_{ij}\notag \\
&-&\frac{\langle\bar{q}q\rangle}{12}\delta_{ij} -\frac{ x^{2}}{192} m_{0}^{2}\langle
\bar{q}q\rangle\Big[1-i\frac{m_q}{6}\!\not\!{x}\Big]\delta_{ij}
\nonumber\\
&+&\frac{i}{3}\Big[\!\not\!{x}\Big(\frac{m_q}{16}\langle
\bar{q}q\rangle-\frac{1}{12}\langle u\Theta^{f}u\rangle\Big)\nonumber\\
&+&\frac{1}{3}\Big(u\cdot x\!\not\!{u}\langle
u\Theta^{f}u\rangle\Big)\Big]\delta_{ij}
\nonumber\\
&-&\frac{ig_s \lambda_{A}^{ij}}{32\pi^{2} x^{2}}
G_{\mu\nu}^{A}\Big(\!\not\!{x}\sigma^{\mu\nu}+\sigma^{\mu\nu}
\!\not\!{x}\Big),
\end{eqnarray}
where $m_{q}$ denotes the light quark mass, $\langle\bar{q}q\rangle$ is the
temperature-dependent light quark condensate, $G_{\mu\nu}^{A}$ is the temperature-dependent external gluon field, 
$\Theta^{f}_{\mu\nu}$ is the fermionic part of the energy momentum tensor and $\lambda_{A}^{ij}$
are the standard Gell-Mann matrices. As is seen, the temperature-dependent condensates are introduced  instead of
 the vacuum saturated condensates (for details see for instance Refs. \cite{Mallik1,Weldon,Mallik}). As also previously mentioned,  the four-velocity vector of the medium $u^{\mu}$  is also introduced 
to restore the Lorentz invariance at finite temperature broken with the choice of
 the thermal rest frame. In the rest frame of the heat bath, the four-velocity vector of the medium is written as  $u^{\mu}=(1,0,0,0)$ which leads to 
 $u^2=1$ and $p\cdot u=p_{0}$. In Eq. (\ref{lightquarkpropagator}), the terms containing the four-quark operators as well as terms with Logarithms are neglected, 
since they  give small contributions to the results. 

By using the above thermal light quark propagator in the coordinate space and performing the Fourier integral to go to momentum space and applying the
 Borel transformations as well as  the continuum subtraction, after lengthy calculations, for the OPE side of the correlation function in the Borel scheme we obtain 

\begin{eqnarray}
\hat{B}\Pi_{\mu\nu}^{OPE}(p_0,T)&=&\Pi_{1}^{OPE}(p_0,T)\!\not\!{p}g_{\mu\nu}\notag \\
&+&\Pi_{2}^{OPE}(p_0,T)g_{\mu\nu}\notag \\
&+&\textsf{other structures},
\end{eqnarray}
where  $\Pi_{1}^{OPE}$ and $\Pi_{2}^{OPE}$ are functions of QCD degrees of freedom as well as the new operators. As examples, we present the explicit forms of these functions  for $\Xi^{*}$ particle in 
 the Appendix.
Not that  we have used the following relation to relate the two-gluon condensate to the gluonic part of the energy-momentum tensor $\Theta^{g}_{\lambda \sigma}$:
\begin{eqnarray}\label{TrGG} 
\langle Tr^c G_{\alpha \beta} G_{\mu \nu}\rangle &=& \frac{1}{24} (g_{\alpha \mu} g_{\beta \nu} -g_{\alpha
\nu} g_{\beta \mu})\langle G^a_{\lambda \sigma} G^{a \lambda \sigma}\rangle \nonumber \\
 &+&\frac{1}{6}\Big[g_{\alpha \mu}g_{\beta \nu} -g_{\alpha \nu} g_{\beta \mu} -2(u_{\alpha} u_{\mu}g_{\beta \nu} \nonumber \\
 &-&u_{\alpha} u_{\nu} g_{\beta \mu} -u_{\beta} u_{\mu}
g_{\alpha \nu} +u_{\beta} u_{\nu} g_{\alpha \mu})\Big]\nonumber \\
&\times&\langle u^{\lambda} {\Theta}^g _{\lambda \sigma} u^{\sigma}\rangle.
\end{eqnarray}

After calculation of the  both  hadronic and OPE sides of the thermal correlation function,
 we match the coefficients of the structures $\!\not\!{p}g_{\mu\nu}$ and $g_{\mu\nu}$ to obtain the  sum rules
\begin{eqnarray}\label{residuesumrule}
-\lambda_D^{2}(T)e^{-m_{D}^2(T)/M^2}=\Pi_{1,D}^{OPE}(p_0,T),
\end{eqnarray}
and
\begin{eqnarray}\label{residuesumrule2}
-\lambda_D^{2}(T)m_{D}(T)e^{-m_{D}^2(T)/M^2}=\Pi_{2,D}^{OPE}(p_0,T),
\end{eqnarray}
by simultaneous calculations of which we get the temperature-dependent mass and residue of the particles under consideration.

\section{Numerical computations and conclusions}

\begin{table}[tbp]
\begin{tabular}{|c|c|}
\hline \hline
   Parameters  &  Values    
           \\
\hline \hline
$p_0^\Delta  $         &  $1.231~GeV$     \\
$p_0  ^{\Sigma^{*}}$        &  $1.383~GeV$     \\
$p_0^{\Xi^{*}}$        &  $1.531~GeV$     \\
$p_0^{  \Omega^{-}}$       &  $1.672~GeV$     \\
$ m_{u}   $          &  $(2.3_{-0.5}^{+0.7})$ $MeV$      \\
$ m_{d}   $          &   $(4.8_{-0.3}^{+0.5})$ $MeV$    \\
$ m_{s}   $          &   $(95\pm5 )$ $MeV$    \\
$ m_{0}^{2}   $          &  $(0.8\pm0.2)$ $GeV^2$         \\
$ \langle0|\overline{u}u|0\rangle = \langle0|\overline{d}d|0\rangle$          &  $-(0.24\pm0.01)^3$ $GeV^3$       \\
$ \langle0|\overline{s}s|0\rangle $          &  $-0.8(0.24\pm0.01)^3$ $GeV^3$       \\
$ {\langle}0\mid \frac{1}{\pi}\alpha_s G^2 \mid 0{\rangle}$          &  $ (0.012\pm0.004)~GeV^4$   \\
\hline \hline
\end{tabular}%
\caption{Input parameters used in calculations \cite{Belyaev,Olive,H.G.Dosch,B.L.Ioffe}. }
\label{tab:Param}
\end{table}

To numerically analyze the sum rules for the masses and residues of $\Delta$, $\Sigma^{*}$, $\Xi^{*}$ and $\Omega^{-}$ decuplet baryons at finite temperature, we use input parameters 
such as the quark masses as well as  the light-quark and gluon condensates in vacuum. They are collected in Table \ref{tab:Param}. Beside these  parameters, 
we need the temperature-dependent quark and gluon condensates as well as the temperature-dependent energy density.
For the quark condensate, we use the following parametrization which reproduces the lattice QCD  and QCD sum 
rules results presented in \cite{Ayala,Bazavov,Cheng1}:
\begin{eqnarray}\label{qbarq}
\langle\bar{q}q\rangle=\frac{\langle0|\bar{q}q|0\rangle}{1+e^{18.10042(1.84692[\frac{1}{GeV^2}] T^{2}+4.99216[\frac{1}{GeV}] T-1)}},
\end{eqnarray}
where $\langle0|\bar{q}q|0\rangle$  is the vacuum light-quark condensate and this function is valid up to a critical temperature $ T_{c}=197~MeV$.

For the gluonic and fermionic parts of the energy density, we get the  parametrization
\begin{eqnarray}\label{tetamumu}
\langle\Theta^{g}_{00}\rangle=\langle\Theta^{f}_{00}\rangle&=&T^{4} e^{(113.867[\frac{1}{GeV^2}] T^{2}-12.190[\frac{1}{GeV}] T)}\nonumber \\
&-&10.141 [\frac{1}{GeV}]T^{5}.
\end{eqnarray}
 obtained using lattice QCD graphics given in \cite{M.Cheng} and valid for $T\geq130~MeV$.

The temperature-dependent gluon condensate obtained using the QCD sum rules predictions and lattice QCD data in \cite{Ayala2} is employed as 
\begin{eqnarray}\label{G2TLattice}
\langle G^{2}\rangle&=&\langle 0|G^{2}|0\rangle\Bigg[1-1.65\Big(\frac{T}{T_{c}}\Big)^{8.735}\nonumber \\
&+&0.04967\Big(\frac{T}{T_{c}}\Big)^{0.7211}\Bigg],
\end{eqnarray}
where $\langle 0|G^{2}|0\rangle$ is the gluon condensate in vacuum. 

The temperature-dependent continuum threshold for decuplet baryons is one of auxiliary parameters that should also be determined.  
For this aim, we use the obtained sum rules for the mass and residue in Eqs.  (\ref{residuesumrule}) and  (\ref{residuesumrule2}) as well as an extra equation obtained from Eq.  (\ref{residuesumrule})
 by applying a derivative with respect to $-\frac{1}{M^2}$ to both sides.  By eliminating the mass and residue from these equations, we get the continuum threshold in terms of temperature, i.e. 
%
\begin{eqnarray}\label{continuumthreshold}
s_{0}(T)=s_{0}\Bigg[1-0.93\Big(\frac{T}{T_{c}}\Big)^{12}\Bigg],
\end{eqnarray}
where  $s_{0}$ is the continuum threshold at $T=0$. This parameter is not totally arbitrary but it depends on the energy of the first excited state with the same quantum numbers as the 
chosen interpolating currents for  decuplet baryons under consideration. The $T$-dependent continuum threshold extrapolates this condition to all temperatures.  For $s_{0}$, we take the interval 
$[m_{D}(0)+0.4]^{2}~GeV^2\leq s_0\leq[m_{D}(0)+0.6]^{2}~GeV^2$ in which the physical quantities show relatively weak dependence on it. By comparison of Eq. (\ref{continuumthreshold}) 
with the expression of the temperature-dependent continuum threshold of  the hyperons   \cite{Azizi2} we see that there is an extra $0.93$ coefficient in the decuplet case.

Finally, we determine the working regions for the Borel mass parameter $M^{2}$.
 For this aim, we require that not only the contributions of the higher states and continuum are adequately  suppressed but also the  portion of
perturbative part  exceeds the non-perturbative contributions and the series of the OPE converge. This leads to the working windows of 
the Borel mass parameter for different decuplet baryons as presented in Table \ref{table3}. 

\begin{table}[h]
\renewcommand{\arraystretch}{1.5}
\addtolength{\arraycolsep}{3pt}
$$
\begin{array}{|c|c|c|c|c|}
\hline \hline
        & M^{2}   \\
\hline
  \mbox{$\Delta$} &1.5~GeV^2\leqslant M^2 \leqslant 3.0~GeV^2 \\
  \hline
  \mbox{$\Sigma^{*}$}& 1.7~GeV^2\leqslant M^2 \leqslant 3.5~GeV^2\\
  \hline
  \mbox{$\Xi^{*}$} &2.0~GeV^2\leqslant M^2 \leqslant 3.8~GeV^2 \\
    \hline
  \mbox{$\Omega^{-}$} &2.2~GeV^2\leqslant M^2 \leqslant 4.0~GeV^2 \\
                    \hline \hline
\end{array}
$$
\caption{ The working regions of $M^{2}$ for $\Delta$, $\Sigma^{*}$, $\Xi^{*}$ and $\Omega^{-}$ decuplet baryons.} \label{fitfunction1}
\renewcommand{\arraystretch}{1}
\addtolength{\arraycolsep}{-1.0pt} \label{table3}
\end{table}

According to the philosophy of the method used, the physical quantities should be practically independent  the auxiliary  parameters $M^{2}$ and $s_{0}$. 
To see how this condition is satisfied for the  decuplet baryons, we plot their mass and residue versus $M^{2}$ for different fixed values of the continuum threshold $s_{0}$ at $T=0$ 
in  figures~ \ref{fig1} and \ref{fig2}.
 From these figures, we see that these quantities depend on both $M^{2}$ and $s_{0}$ very weakly  in their working intervals. 

The final task is to investigate the variations of the mass and residue of  $\Delta$, $\Sigma^{*}$, $\Xi^{*}$ and $\Omega^{-}$  baryons with respect to temperature and compare 
the obtained results on the behavior of these quantities with respect temperature  with our previous works for nucleon and hyperons  \cite{Azizi1, Azizi2} as well as existing 
predictions in the literature \cite{Rincon,Bedaque,Xu}. For this purpose,
 we plot these quantities as a function of temperature in figures  \ref{fig3} and \ref{fig4}.  These figures indicate that the mass and residue of  $\Delta$, $\Sigma^{*}$, $\Xi^{*}$ 
and $\Omega^{-}$ decuplet baryons remain approximately unchanged up to $T\cong0.15~ GeV$, however, after this point, they start to diminish rapidly by increasing the temperature. 
The main reason behind such behavior  is the variation of the $T$-dependent continuum threshold with respect to temperature  presented in Eq. (\ref{continuumthreshold}) and calculated, for the first time in the present
study, in Decuplet channel. Our analyses show that this quantity
does not change up to $T\cong0.15~GeV $, beyond which it drastically falls. At the critical or deconfinement temperature, the continuum threshold falls with amount of $93\%$.
As we already mentioned the continuum threshold separates the ground state from the higher states and continuum. It appears in the upper limits of many integrals in the sum rules obtained for the masses and residues (see the OPE expressions in the Appendix) where their lower limits are roughly zero because of the light quark masses. When temperature is increased the region of integration gets smaller. 
The  temperature-dependent quark condensate that gives higher contribution after the perturbative part also  shows similar behavior 
and kills many terms in the expressions of the sum rules as it approaches to zero near to the critical temperature. 
Such variations of the parameters of the hadrons under consideration near to the critical temperature  can be considered as a sign of transition to the quark-gluon-plasma (QGP) as the new phase of the matter.

The average results on 
the masses obtained in the limit $T\rightarrow0$ are  $m_{\Delta}(0)=1.239\pm0.148~GeV$, $m_{\Sigma^{*}}(0)=1.394\pm0.167~GeV$, $m_{\Xi^{*}}(0)=1.525\pm0.183~GeV$ 
and  $m_{\Omega^{-}}(0)=1.693\pm0.203~GeV$ which, within the errors, are in  good consistencies with the experimental data \cite{Olive}. We also  obtain the average values
 $\lambda_{\Delta}(0)=0.038\pm0.010~GeV^3$, $\lambda_{\Sigma^{*}}(0)=0.043\pm0.012~GeV^3$, $\lambda_{\Xi^{*}}(0)=0.053\pm0.014~GeV^3$ 
and  $\lambda_{\Omega^{-}}(0)=0.068\pm0.019~GeV^3$ for
the residues at zero temperature which are also  consistent with the results of Ref. \cite{Lee2}. Looking at the values of zero-temperature masses and residues we see that there is a considerably large
 SU(3) flavor  symmetry breaking effect from $\Delta$ (consist of up and down quarks) and $\Omega^{-}$ (consist of only s quark). These violations are with amounts of $27\%$ and $44\%$ for the mass  and
residue, respectively. However, when considering the thermal  behavior of the mass and residue of decuplet baryons we can not detect any SU(3) flavor violation effects and all members demonstrate roughly the same trend 
with respect to temperature.
 Near to de-confinement or critical temperature, the 
masses of $\Delta$, $\Sigma^{*}$, $\Xi^{*}$ and $\Omega^{-}$ decuplet baryons fall with amount of roughly $80\%$,  while the residues overall reduce 
with amount of  $35\%$ compared to their vacuum values. Though the zero-temperature values of the mass and residue of the decuplet baryons are very different than those 
of the octet baryons \cite{Azizi1, Azizi2}, the behaviors of the mass and residue of the decuplet baryons
 in terms of temperature obtained in the present work are similar to those of the nucleon and hyperons  \cite{Azizi1, Azizi2} and we do not see the effect of the spin on the
thermal properties of baryons with the same quark contents.  The variations of the masses versus temperature for decuplet baryons obtained in 
the present study  are also in agreement with those of  Ref. \cite{Rincon,Xu}, i.e., in all of these studies the masses reduce rapidly with respect to $T$ near to the critical temperature.

In conclusion, we investigated the mass and residue of $\Delta$, $\Sigma^{*}$, $\Xi^{*}$ and $\Omega^{-}$ decuplet baryons at finite temperature in the framework of thermal QCD sum rules to acquire sum rules
for these quantities in terms of different QCD degrees of freedom and the new operators coming up at finite temperature. We found the fit function for the temperature-dependent continuum threshold in terms of 
vacuum threshold. By fixing the auxiliary parameters,  $M^{2}$ and  $s_{0}$ and using the temperature-dependent quark and gluon condensates as well as the thermal average of the energy density
obtained via lattice QCD and QCD sum rules, we analyzed the behaviors of the masses and residues versus temperature.
 We observed that the mass and residue of the decuplet baryons remain unchanged with the variations of temperature up to $T\cong0.15~ GeV$, after this point they decrease rapidly such that near to the 
  critical  temperature the masses and residues reach to roughly $20\%$ and $65\%$ of their vacuum values, respectively.  The melting of the baryons near to the critical or deconfinement temperature may be considered as a sign of transition to 
  QGP phase that are searched for at different Colliders.
We also extracted the numerical values of the masses and residues at $T\rightarrow0$ limit  which are in good consistencies with the existing experimental data as well 
as other predictions in the literature. 

Our results may be used in analyses of the results of the future heavy ion collision experiments. The $T$-dependent quantities considered in the present work, especially
the temperature dependent residues, can be used in the study of the strong couplings of the particles under consideration with other hadrons  as well as their electromagnetic properties
and radiative decays at finite temperature. The residue is the main input parameter  in such studies.


\label{sec:results}

\section*{ACKNOWLEDGEMENTS}

This work has been partly supported by the Scientific and Technological
Research Council of Turkey (TUBITAK) under the national postdoctoral research scholarship program 2218.



\begin{widetext}

\appendix*

\section{OPE expressions for $\Xi^{*}$}
\renewcommand{\theequation}{\Alph{section}.\arabic{equation}}

\label{sec:AppB}
%
In this appendix, as examples, we present the functions $\Pi_{1}^{OPE}(p_0,T)$ and  $\Pi_{2}^{OPE}(p_0,T)$  for $\Xi^{*}$  baryon which are obtained as 

\begin{eqnarray}\label{borelpi1}
\Pi_{1}^{OPE}(p_0,T)&=&\frac{1}{160\pi^4}\int^{s_0(T)}_{(2m_{s}+m_{u})^2}ds \exp\Big(-\frac{s}{M^2}\Big)s^2\nonumber\\
&+&\frac{\langle\bar{q}q\rangle}{48\pi^2}\Bigg[m_{0}^2\Big(8m_{s}-m_{u}\Big)+\Big(4m_{u}-16m_{s}\Big)\int^{s_0(T)}_{(2m_{s}+m_{u})^2}ds \exp\Big(-\frac{s}{M^2}\Big)\Bigg]\nonumber\\
&+&\frac{\langle\bar{s}s\rangle}{24\pi^2}\Bigg[m_{0}^2\Big(3m_{s}+4m_{u}\Big)-4\Big(m_{s}+2m_{u}\Big)\int^{s_0(T)}_{(2m_{s}+m_{u})^2}ds \exp\Big(-\frac{s}{M^2}\Big)\Bigg]\nonumber\\
&-&\frac{\langle u\Theta^{f} u \rangle}{9\pi^2}\Bigg[4p_{0}^2-\int^{s_0(T)}_{(2m_{s}+m_{u})^2}ds\exp\Big(-\frac{s}{M^2}\Big)\Bigg]+\frac{\alpha_{s} \langle u\Theta^{g} u \rangle}{9\pi^3} \int^{s_0(T)}_{(2m_{s}+m_{u})^2}ds\exp\Big(-\frac{s}{M^2}\Big)\nonumber\\
&+&\frac{5\langle\alpha_{s}G^2\rangle}{144\pi^3}\int^{s_0(T)}_{(2m_{s}+m_{u})^2}ds\exp\Big(-\frac{s}{M^2}\Big)-\langle\bar{s}s\rangle^2 \Bigg[\frac{(2m_{0}^2 - 4M^2)}{9M^2}\Bigg]-\langle\bar{q}q\rangle \langle\bar{s}s\rangle \Bigg[\frac{(4m_{0}^2 - 8M^2)}{9M^2}\Bigg]\nonumber \\
&+&\frac{\langle\bar{q}q\rangle\langle u\Theta^{f} u \rangle}{81M^6}\Bigg\{2m_{0}^2  m_{u}(-7M^2 - 8p_{0}^2)+48M^2\Bigg[ m_{s}M^2+m_{u}\Big(M^2 +2p_{0}^2\Big)\Bigg]\Bigg\}\nonumber\\
&+&\frac{\langle\bar{s}s\rangle\langle u\Theta^{f} u \rangle}{81M^6}\Bigg\{4m_{0}^2  m_{s}(-7M^2 - 8p_{0}^2)+48M^2\Bigg[ 3m_{s}M^2+m_{u}M^2+4m_{s}p_{0}^2\Bigg]\Bigg\}\nonumber\\
&+&\Big[\frac{\alpha_{s}\langle\bar{q}q\rangle\langle u\Theta^{g} u \rangle}{108\pi M^4}m_{u}+\frac{\alpha_{s}\langle\bar{s}s\rangle\langle u\Theta^{g} u \rangle}{54\pi M^4}m_{s}\Big] \Big(3m_{0}^2-8M^2\Big)\nonumber\\
&-&\Big[\frac{\langle\bar{q}q\rangle\langle\alpha_{s}G^2\rangle}{432\pi M^4}m_{u}+\frac{\langle\bar{s}s\rangle\langle\alpha_{s}G^2\rangle}{216\pi M^4}m_{s}\Big]\Big(3m_{0}^2 - 20M^2\Big)+\frac{2\alpha_{s} \langle u\Theta^{f} u \rangle \langle u\Theta^{g} u \rangle}{27\pi M^4}(M^2 +p_{0}^2)\nonumber\\
&-&\frac{\langle\alpha_{s}G^2\rangle \langle u\Theta^{f} u \rangle}{27\pi M^4}(3M^2 -4p_{0}^2)-\frac{2\langle u\Theta^{f} u \rangle^2}{9 M^2}(5 + 2t + 5t^2),
\end{eqnarray}
\begin{eqnarray}\label{borelpi2}
\Pi_{2}^{OPE}(p_0,T)&=&\frac{(2m_{s} + m_{u})}{64\pi^4}\int^{s_0(T)}_{(2m_{s}+m_{u})^2}ds \exp\Big(-\frac{s}{M^2}\Big)s^2\nonumber\\
&+&\frac{\langle\bar{q}q\rangle+2\langle\bar{s}s\rangle}{72\pi^2} \Bigg[ \int^{s_0(T)}_{(2m_{s}+m_{u})^2}(3m_{0}^2-8s) ds\exp\Big(-\frac{s}{M^2}\Big)\Bigg]\nonumber\\
&+&\frac{2\langle u\Theta^{f} u \rangle}{9\pi^2} (2 m_{s}+ m_{u}) \Bigg[ p_{0}^2+\int^{s_0(T)}_{(2m_{s}+m_{u})^2} ds\exp\Big(-\frac{s}{M^2}\Big)\Bigg]\nonumber\\
&+&\frac{\alpha_{s} \langle u\Theta^{g} u \rangle}{36\pi^3}(2 m_{s}+ m_{u}) \int^{s_0(T)}_{(2m_{s}+m_{u})^2} ds\exp\Big(-\frac{s}{M^2}\Big)\nonumber\\
&+&\frac{\langle\alpha_{s}G^2\rangle}{48\pi^3}(2 m_{s}+ m_{u}) \int^{s_0(T)}_{(2m_{s}+m_{u})^2} ds\exp\Big(-\frac{s}{M^2}\Big)\nonumber\\
&+&\langle\bar{s}s\rangle^2 \Bigg[\frac{-5m_{0}^2 m_{s}+ 18m_{u}M^2}{27M^2}\Bigg]+\langle\bar{q}q\rangle \langle\bar{s}s\rangle \Bigg[\frac{-5m_{0}^2 (m_{s}+m_{u})+36m_{s}M^2}{27M^2}\Bigg]\nonumber \\
&+&\frac{4\langle\bar{q}q\rangle\langle u\Theta^{f} u \rangle}{27M^4} \Big[-4M^2(2 M^2+p_{0}^2)+m_{0}^2(3M^2+p_{0}^2) \Big]\nonumber \\
&+&\frac{8\langle\bar{s}s\rangle\langle u\Theta^{f} u \rangle}{27M^4}\Big[-4M^2(2 M^2+p_{0}^2)+m_{0}^2(3M^2+p_{0}^2) \Big]-\frac{2\alpha_{s}\langle\bar{q}q\rangle\langle u\Theta^{g} u \rangle}{27\pi M^2}(m_{0}^2-2M^2)\nonumber\\
&-&\frac{4\alpha_{s}\langle\bar{s}s\rangle\langle u\Theta^{g} u \rangle}{27\pi M^2}(m_{0}^2-2M^2)+\frac{\langle\bar{q}q\rangle\langle\alpha_{s}G^2\rangle}{54\pi M^2}(m_{0}^2-4M^2)+\frac{\langle\bar{s}s\rangle\langle\alpha_{s}G^2\rangle}{27\pi M^2}(m_{0}^2-4M^2) \nonumber\\
&+&\frac{16\langle u\Theta^{f} u \rangle^2}{81 M^4}(2m_{s}+m_{u})(3M^2 + 14p_{0}^2).
\end{eqnarray}

\end{widetext}

\begin{widetext}

\begin{figure}[ht]
\begin{center}
\subfigure[]{\includegraphics[width=8cm]{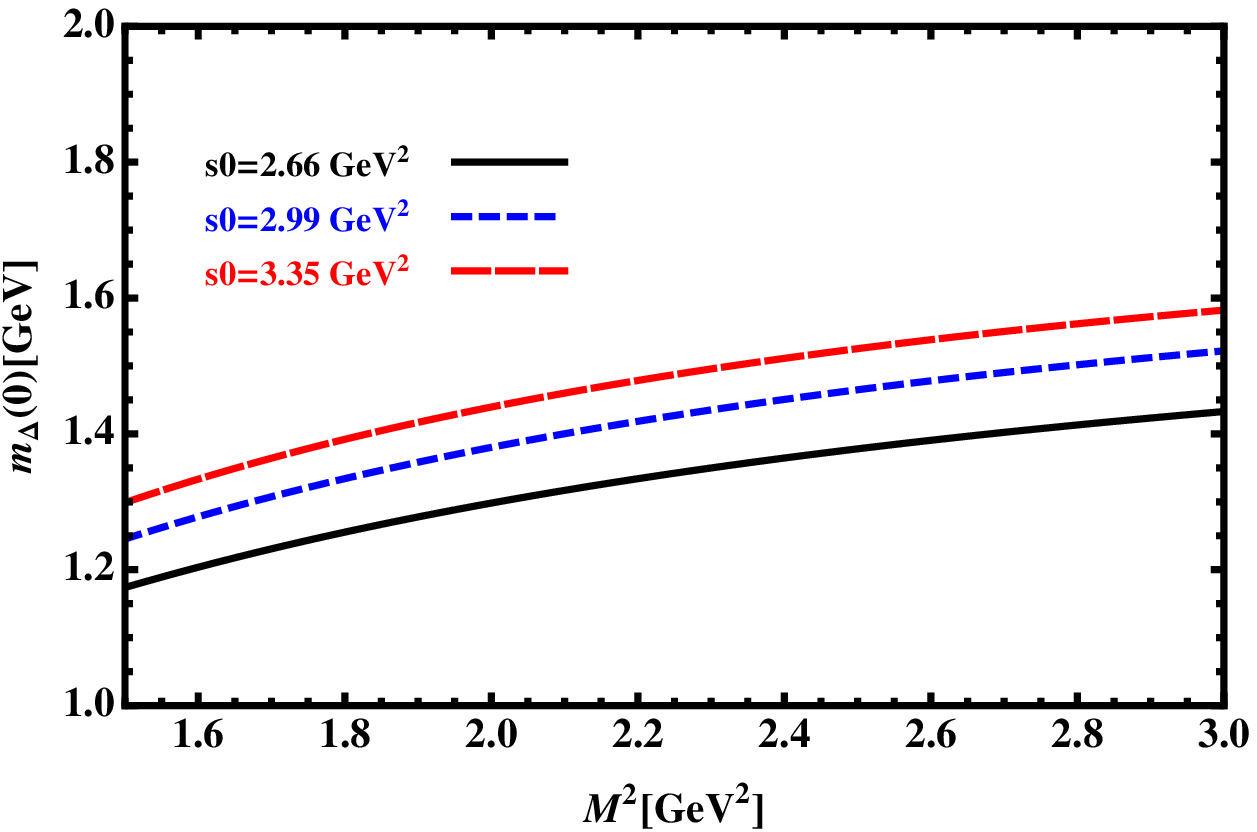}}
\subfigure[]{\includegraphics[width=8cm]{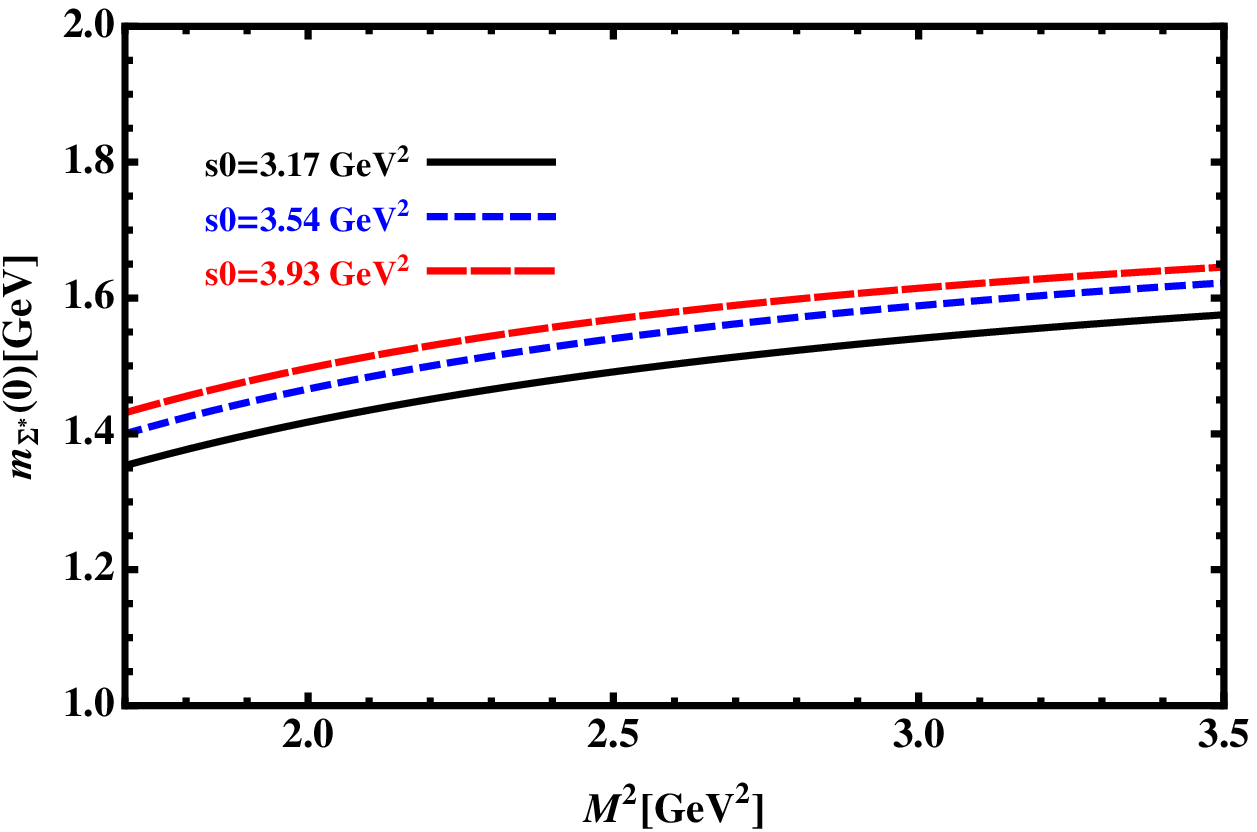}}
\subfigure[]{\includegraphics[width=8cm]{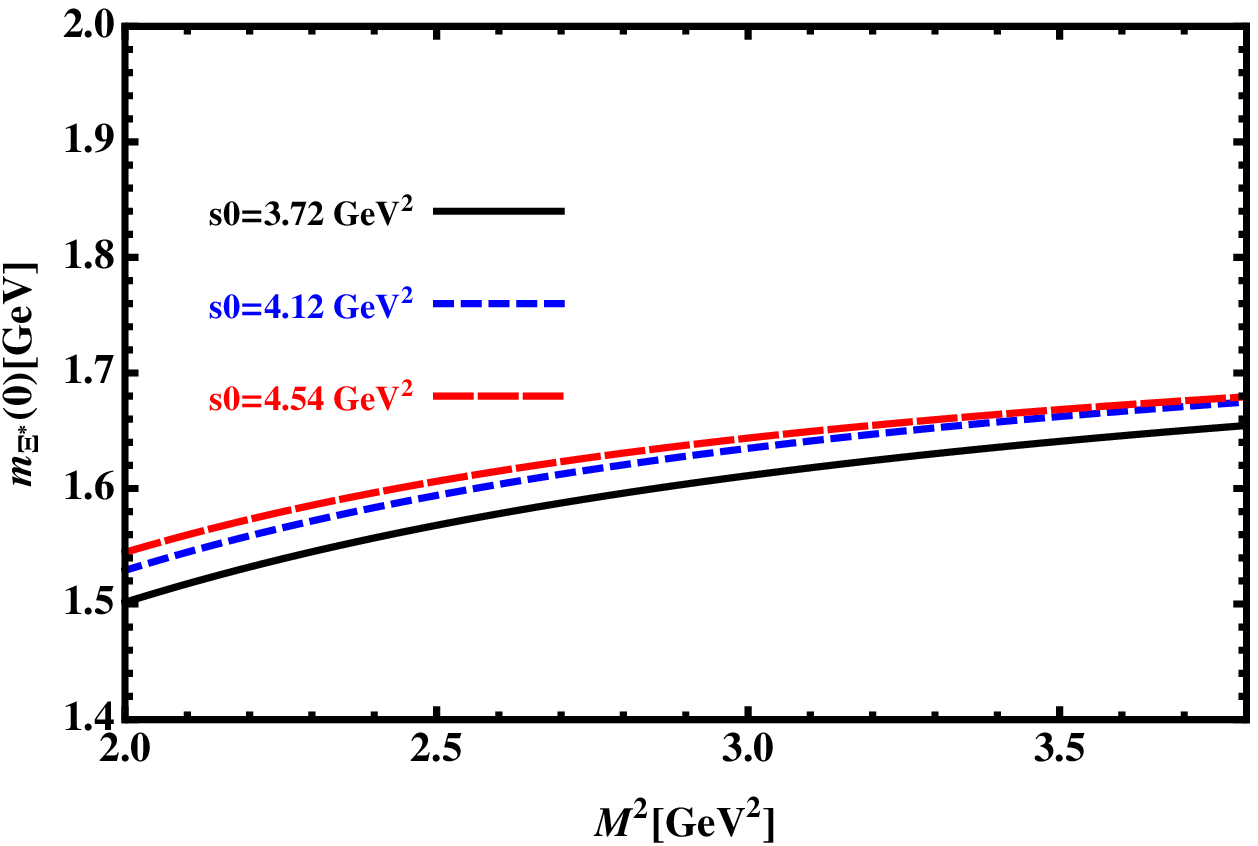}}
\subfigure[]{\includegraphics[width=8cm]{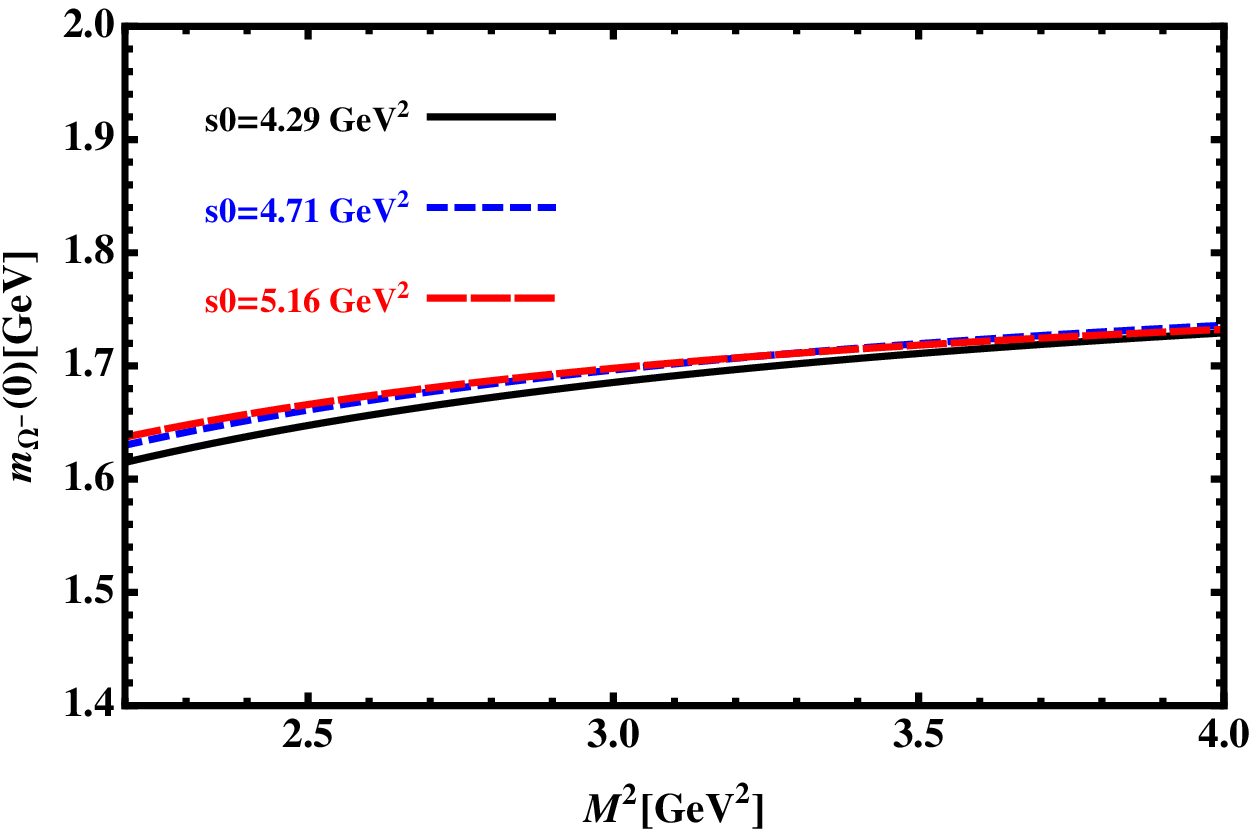}}
\end{center}
\caption{(a) The mass of the  $\Delta$  baryon as a function of $M^2$ for different 
fixed values of $s_0$ at $T=0$. (b)  The same as (a) but for $\Sigma^{*}$  baryon. (c)  The same as (a) but for  $\Xi^{*}$  baryon. (d)  The same as (a) but for  $\Omega^{-}$  baryon.} \label{fig1}
\end{figure}
\begin{figure}[ht]
\begin{center}
\subfigure[]{\includegraphics[width=8cm]{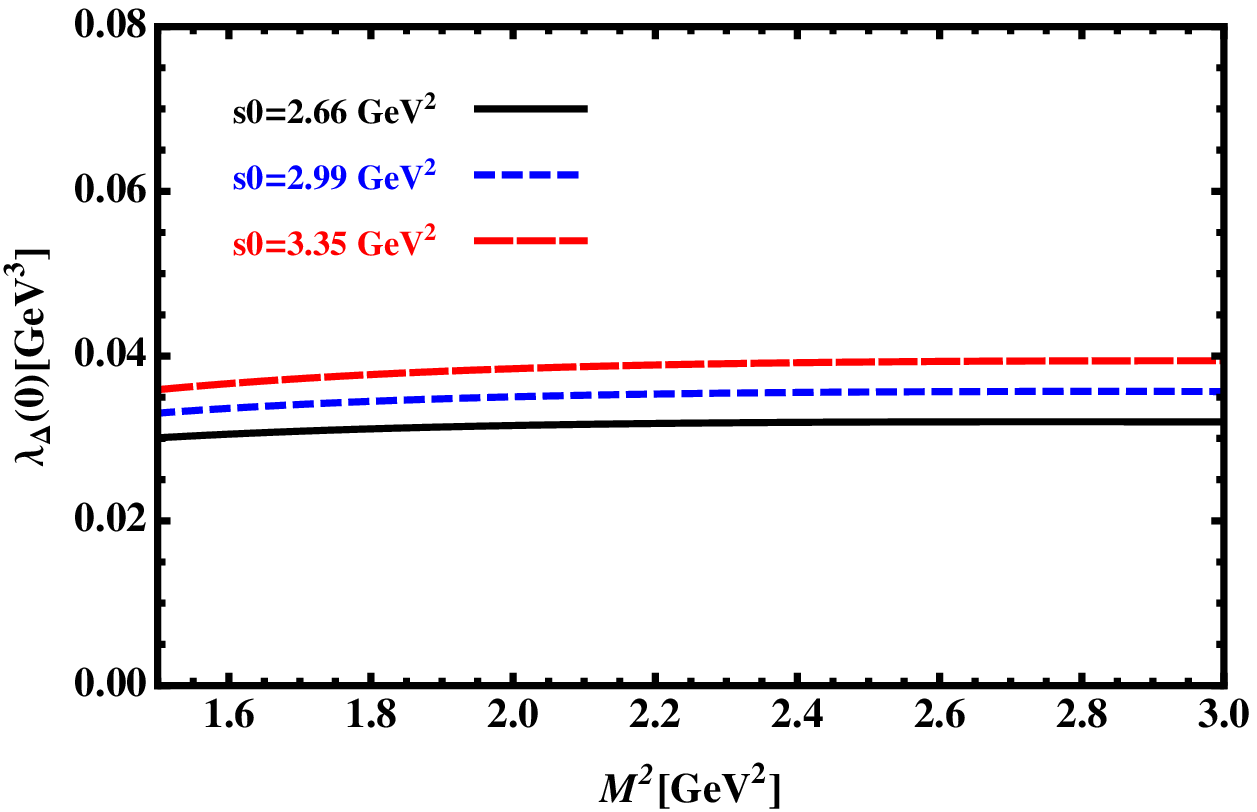}}
\subfigure[]{\includegraphics[width=8cm]{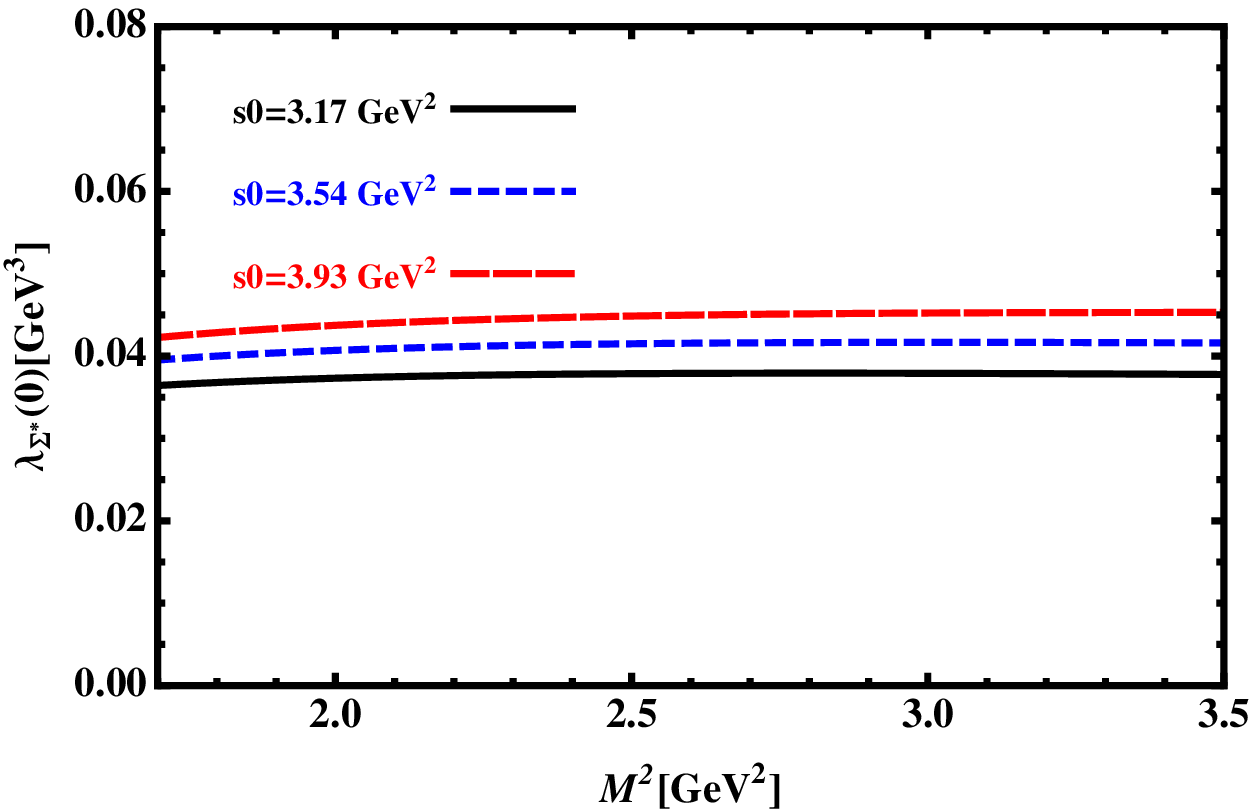}}
\subfigure[]{\includegraphics[width=8cm]{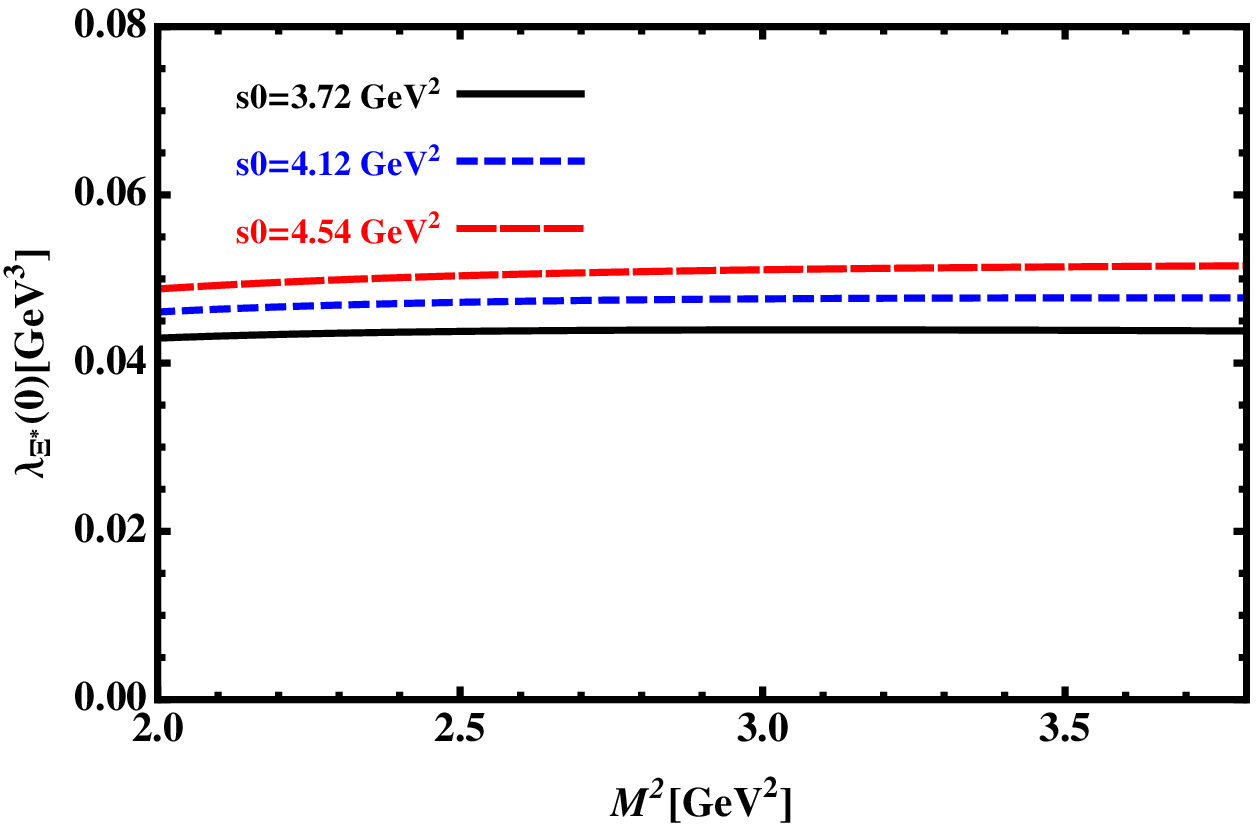}}
\subfigure[]{\includegraphics[width=8cm]{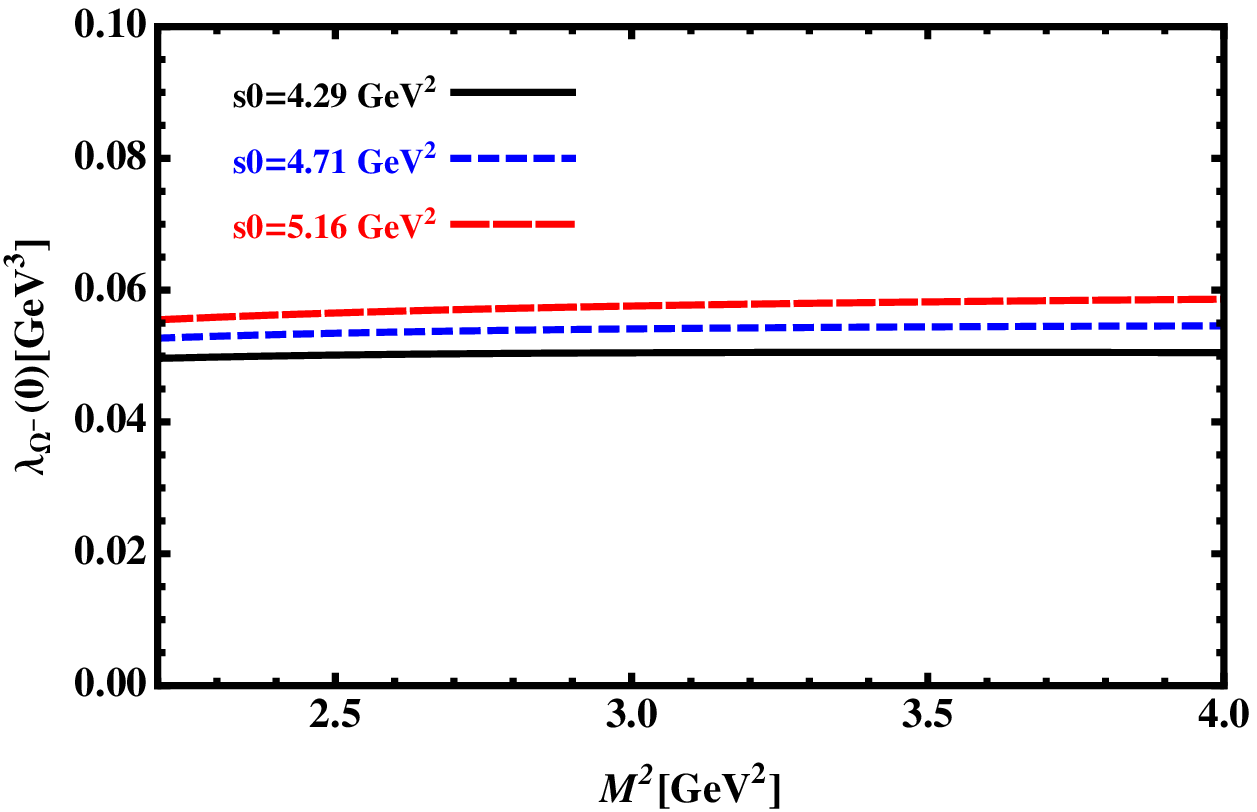}}
\end{center}
\caption{(a) The residue of the $\Delta$  baryon as a function of $M^2$ for different 
fixed values of $s_0$ at $T=0$. (b)  The same as (a) but for $\Sigma^{*}$  baryon. (c)  The same as (a) but for  $\Xi^{*}$  baryon. 
(d)  The same as (a) but for  $\Omega^{-}$  baryon.} \label{fig2}
\end{figure}
\begin{figure}[ht]
\begin{center}
\subfigure[]{\includegraphics[width=8cm]{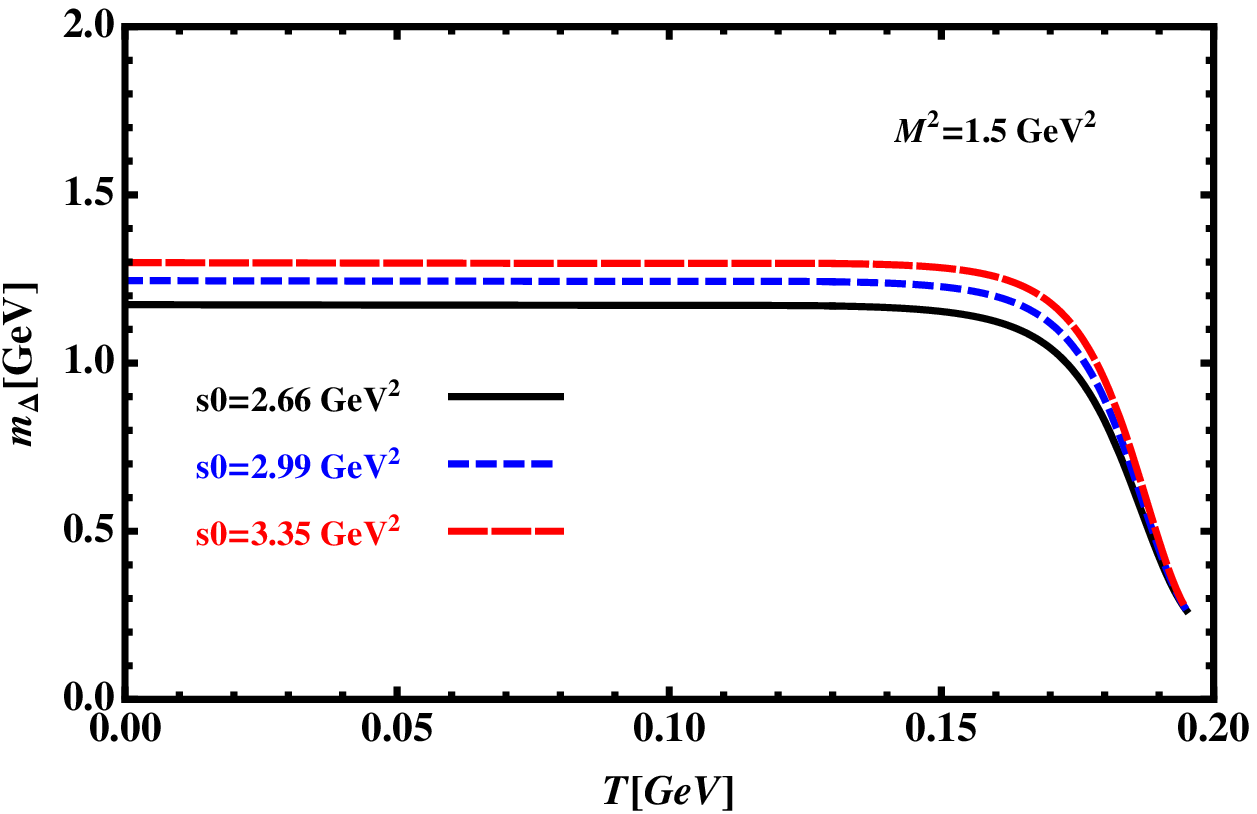}}
\subfigure[]{\includegraphics[width=8cm]{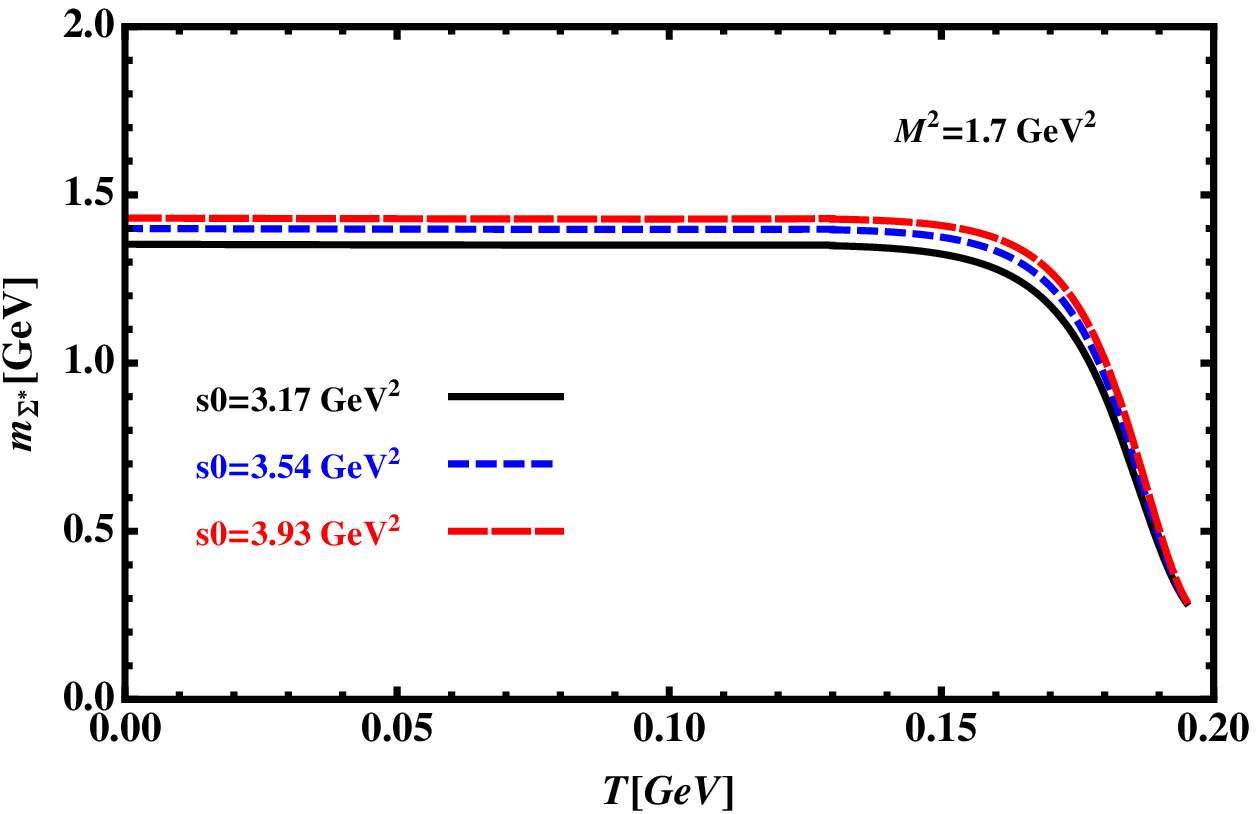}}
\subfigure[]{\includegraphics[width=8cm]{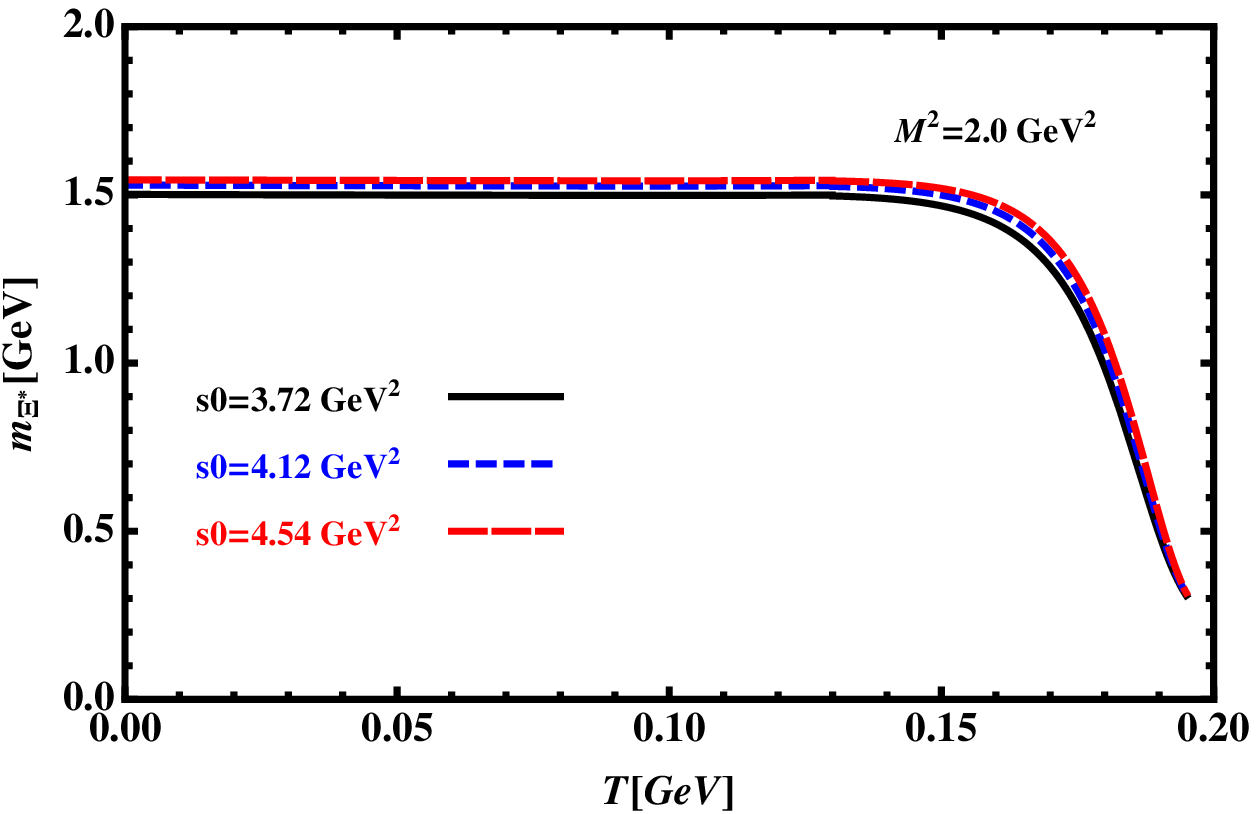}}
\subfigure[]{\includegraphics[width=8cm]{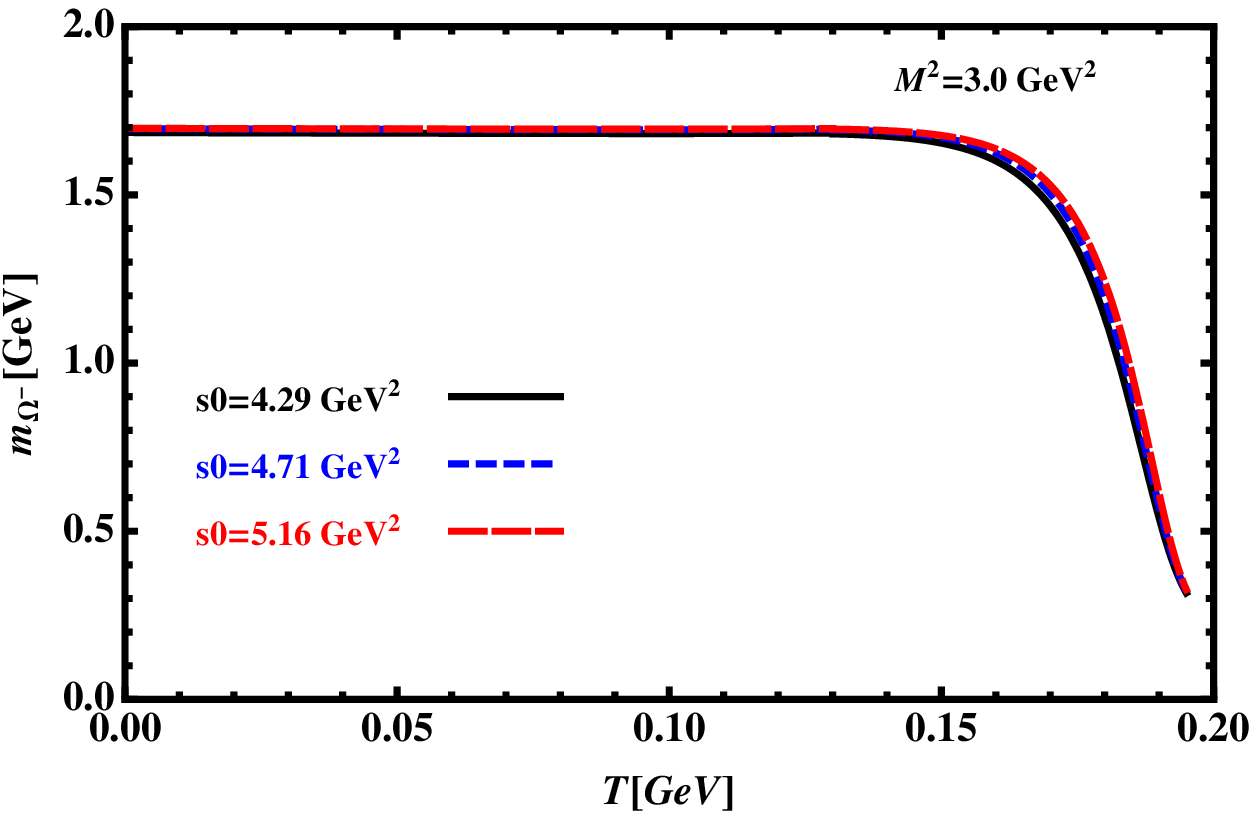}}
\end{center}
\caption{(a) The mass of the  $\Delta$  baryon as a function of temperature at   fixed values of $s_0$ and $M^{2}$. (b) The same as (a) but for $\Sigma^{*}$ 
 baryon. (c) The same as (a) but for $\Xi^{*}$  baryon. (d) The same as (a) but for $\Omega^{-}$  baryon.} \label{fig3}
\end{figure}
\begin{figure}[ht]
\begin{center}
\subfigure[]{\includegraphics[width=8cm]{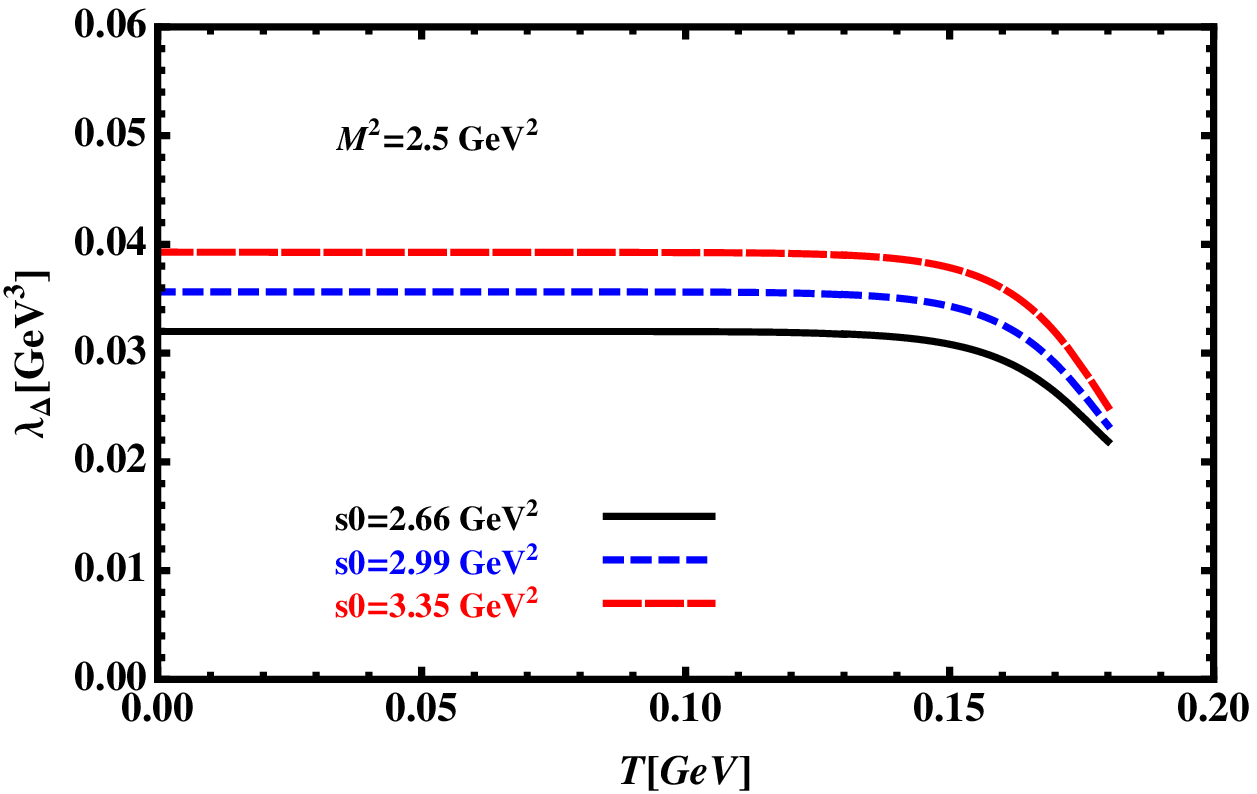}}
\subfigure[]{\includegraphics[width=8cm]{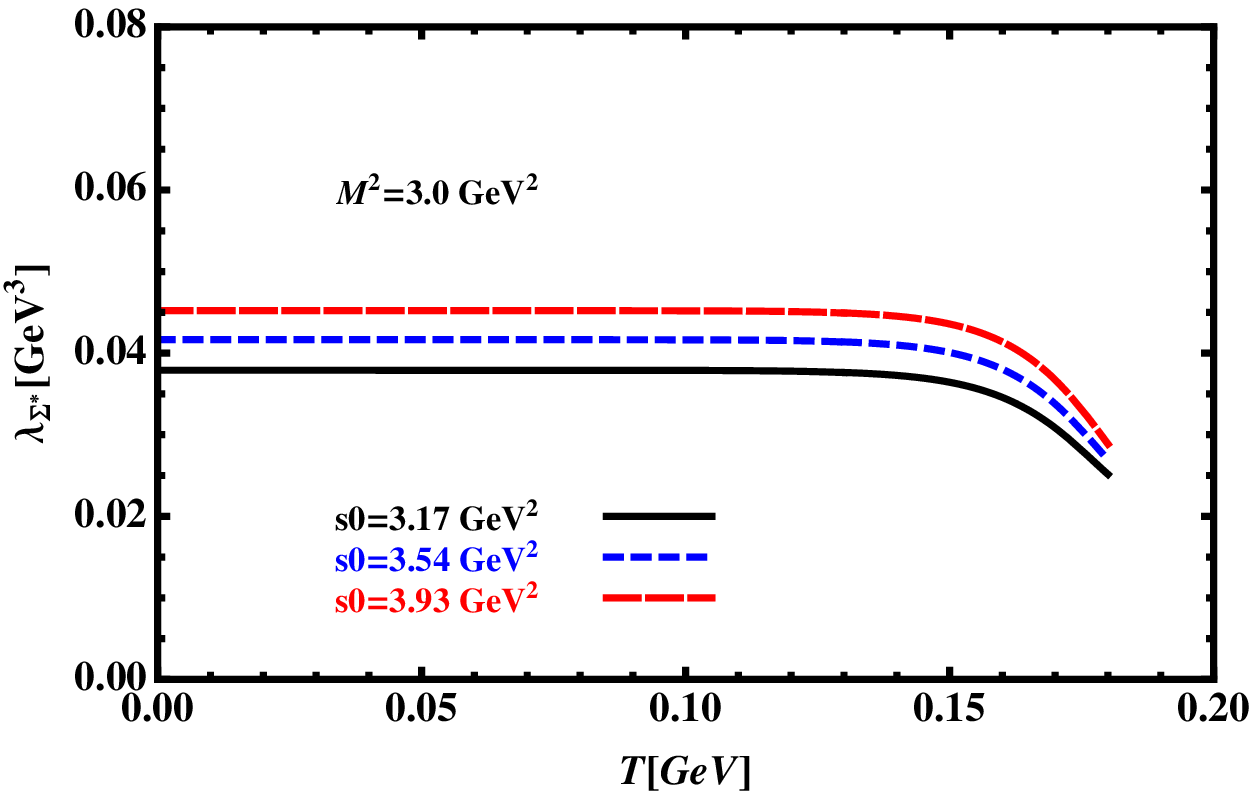}}
\subfigure[]{\includegraphics[width=8cm]{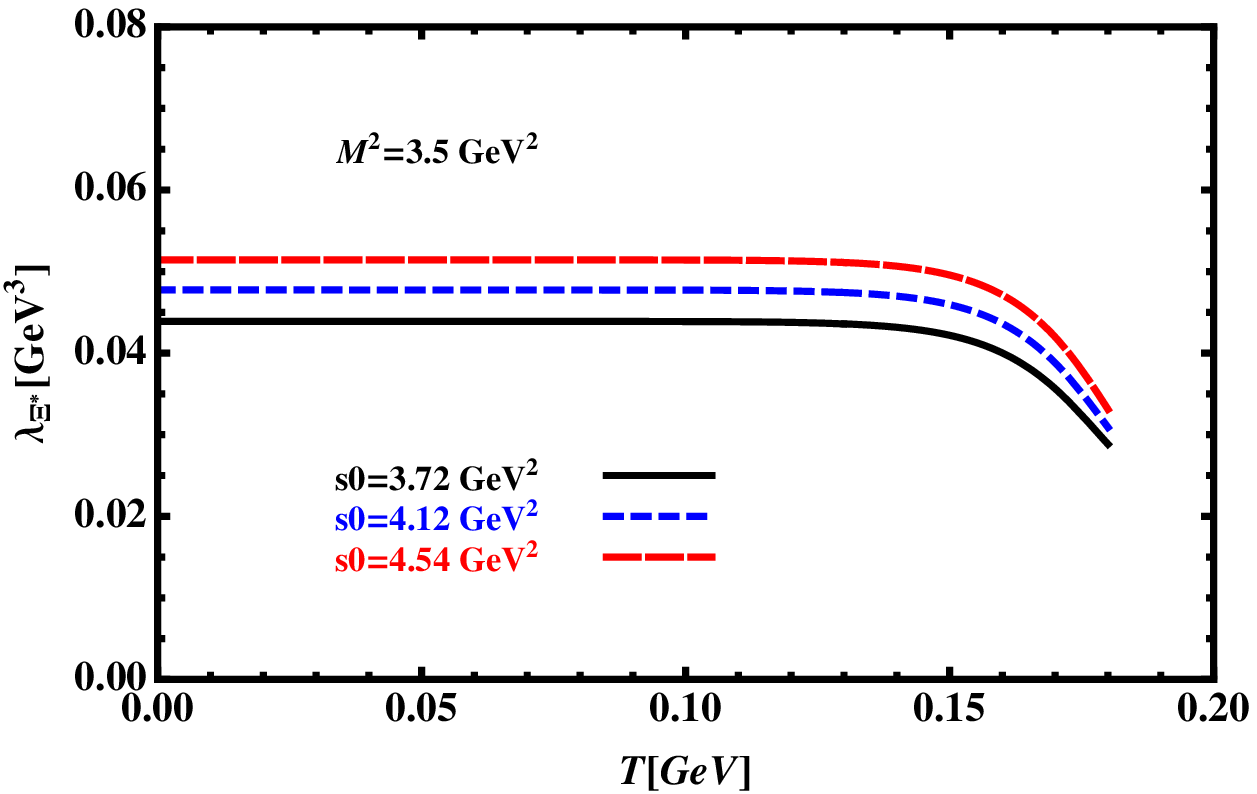}}
\subfigure[]{\includegraphics[width=8cm]{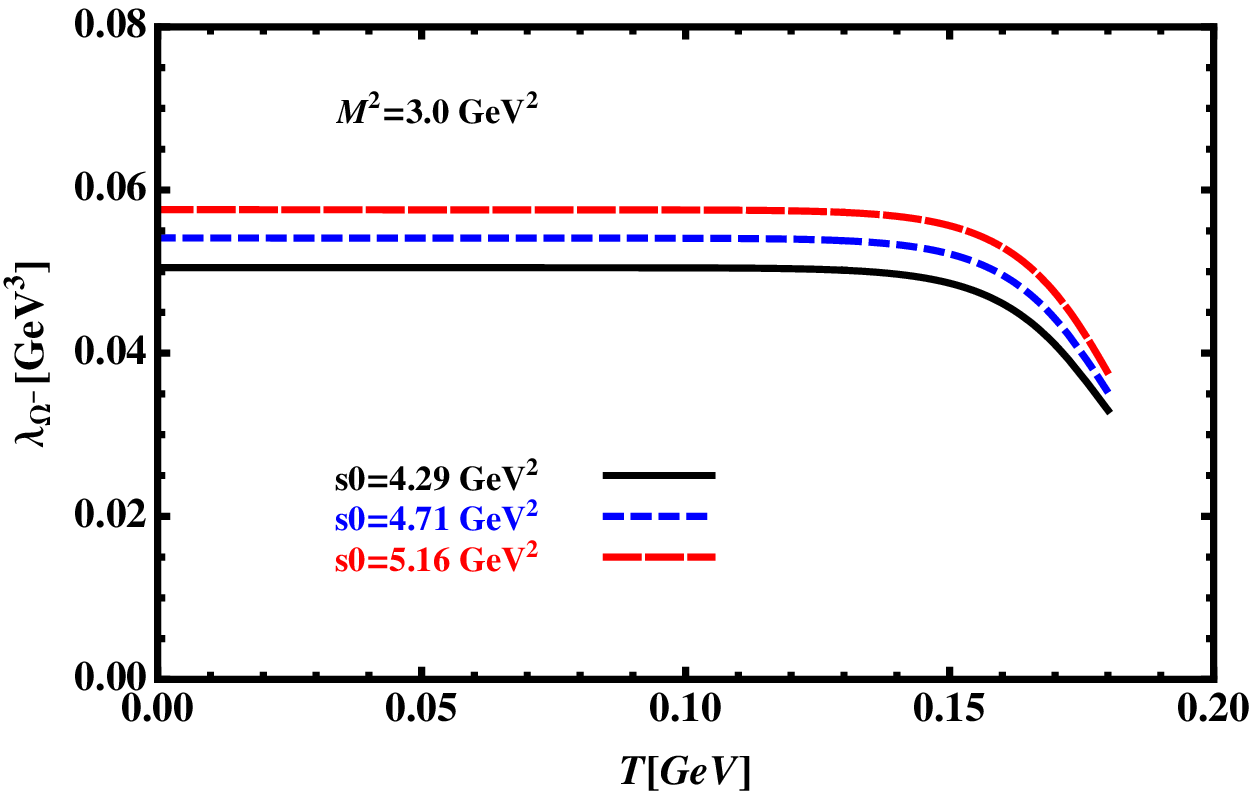}}
\end{center}
\caption{(a) The residue of the  $\Delta$  baryon as a function of temperature at  fixed values of $s_0$ and $M^{2}$.
 (b) The same as (a) but for $\Sigma^{*}$  baryon.  (c) The same as (a) but for $\Xi^{*}$  baryon. 
(d) The same as (a) but for $\Omega^{-}$  baryon.} \label{fig4}
\end{figure}
\end{widetext}

\end{document}